\documentstyle[12pt]{article}

\begin{document}
\title{\vskip -2cm
\Large\bf  ~\\ Non-commutative black holes\\ in $D$ dimensions
}
\vskip .5 cm
\author{~\\C.~Klim\v{c}\'{\i}k\thanks{e-mail: presov @
cspuni12.bitnet} $^1$, P.~Koln\'{\i}k\thanks{e-mail: kolnik
@ cspuni12.bitnet} $^1$, and
A.~Pompo\v{s} \thanks{e-mail: besa @
cspuni12.bitnet}
$^2$\\
{}~\\
\small $^1$ Theory Division of the Nuclear Centre,
Charles University,\\
\small  V Hole\v{s}ovi\v{c}k\'ach 2, 180 00
Prague 8, Czech Republic\\
\small $^2$ Department of Theoretical Physics,
Charles University,\\
\small V Hole\v{s}ovi\v{c}k\'ach 2, 180 00
Prague 8, Czech Republic
\date{\small April, 1994}}
\begin{titlepage}
\maketitle
\vspace*{-11cm}

\leftline{{\it Nuclear Centre of Charles University, Prague}\hfill
PRA-HEP-94/5}
\leftline{\it Institute of Physics, Prague}
\vspace*{9cm}
\begin{abstract}
{\small Recently introduced classical theory of gravity in
non-commutative geometry is studied. The most general (four
parametric) family of $D$ dimensional static spherically
symmetric spacetimes is identified and its properties are
studied in detail. For wide class of the choices of parameters,
the corresponding spacetimes have the structure of
asymptotically flat black holes with a smooth event horizon
hiding the curvature singularity. A specific attention is
devoted to the behavior of components of the metric in
non-commutative direction, which are interpreted as the black
hole hair. }
\end{abstract}
\centerline{PACS index: 04.20, 04.30}
\end{titlepage}

\def\be{\begin{equation}}
\def\ee{\end{equation}}
\def\bea{\begin{eqnarray}}
\def\eea{\end{eqnarray}}

\section{Introduction}

General relativity is, in a sense, the most interesting field
theory because the propagating field -- the metric tensor --
encodes the geometrical properties of the space-time in which
all other fields propagate. Thus, the gravity field couples to
all remaining matter and in this sense it is universal. It was
always very tempting to give a geometrical interpretation also
to other physically relevant fields, like electromagnetic or
Yang-Mills potentials. The oldest such scenario is the
Kaluza-Klein one in which the components of the metric tensor
corresponding to N additional coordinates of the 4+N-dimensional
space-time play the role of the matter fields
\cite{Kaluza,Klein,deWitt}.
There are many variants of the Kaluza-Klein approach but all of
them have some common features.  Namely, the higher dimensional
space is ``real" and one has to look for realistic dynamical
compactification which would lead to vacua with a very tiny size
in the N extra dimensions. Thus, the study of the
$D$-dimensional systems may be something far more than an
academic exercise.

Among more modern geometrical theories of matter the important
role is played by string theory \cite{GSchW}. The dynamics of
the massless modes of the string is governed by an effective
action which possesses the reparametrization invariance and
describes the interaction of the metric tensor with the axion
and dilaton fields.  The important lesson to be learnt from this
consists in emergence of new matter field multiplets (i.e.~axion
and dilaton) in the theory, reflecting its underlying
geometrical structure.  In this sense, those new fields can be
understood as being of the geometrical origin.

Recently, some activity was devoted to the formulation of
general relativity in the framework of non-commutative geometry
of A.~Connes
\cite{Chff1,KPS}. The fundamental algebra of the non-commutative
geometry was chosen to be the algebra of two by two diagonal
matrices with $C^{\infty}$ functions on some $D$-dimensional
manifold as the entries.  Speaking more intuitively, the
space-time consisted of two smooth manifolds--the sheets and it
had a component measuring the distance of the sheets (for the
details see \cite{Chff1,KPS}). Thus, apart from the standard
general relativity metric, there were components taking into
account the relation between two sheets. Indeed those components
appeared in the final action as the ``matter" fields from the
point of view of the standard general relativity. In this way we
may try to give geometrical meaning to various matter fields,
using an appropriate non-commutative geometry setting.

In the paper by Chamseddine, Felder and Fr\"ohlich \cite{Chff1},
the geometrical interpretation was given to the massless scalar
field coupled to general relativity. On the other hand, in the
work by Klim\v{c}\'{\i}k, Pompo\v{s} and Sou\v{c}ek \cite{KPS}
the evaluation of the non-commutative Einstein-Hilbert action
resulted in somewhat exotic coupling of the vector field to
standard general relativity Lagrangian:
\begin{equation}
I=\int_X
d^D x\sqrt{g}[2R+
Q^{\alpha\beta\gamma\delta}
(V)\nabla_{\alpha}V_{\beta}\nabla_{\gamma}
V_{\delta}],
\end{equation}
where
$$
Q^{\alpha\beta\gamma\delta}(V)=4\frac{1}{(V^2)^3}\Big(V^{\alpha}
V^{\beta}V^{\gamma}V^{\delta}-
g^{\alpha\beta}V^{\gamma}V^{\delta}V^2
- g^{\gamma\beta}V^{\alpha}V^{\delta}V^2\Big),
$$
and $\nabla_\mu$ is covariant derivative.
We have written the corresponding action in $D$ dimensions
because the main
point of this paper consists in studying the properties of the
system in
dependence on $D$. In four dimensions we have found the
solutions of the
model which the black hole structure with the $V^\alpha$
field playing the role of
hair \cite{KKP1}. It turned out that while the metric was the
standard black-hole-like one with the smooth horizon,
singularity at the origin and vanishing curvature in the
asymptotical region the hair was less ``healthy". Indeed there
was a critical radius (which could lie below the horizon) under
which the hair became imaginary! This is very peculiar
singularity of the hair because otherwise nothing happens at
that point - the curvature is smooth and bounded as well as the
hair is.

Considering the dimension of the spacetime as a parameter of
the physical theories has brought about many important insights
about the dynamics of physical systems \cite{Kli}. The
higher-dimensional
theories can be relevant in the Kaluza-Klein scenario and
lower-dimensional ones describe dynamics of some particular
subsets of degrees of freedom of four dimensional models.
It is therefore natural to study the dimension dependence
of our non-commutative theories.

In the present case we were very much interested what is going
to be the destiny of the ``imaginary hair" pathology in
arbitrary dimension. We have found, by solving quite a
complicated system of equations, that due to specific structure
of the field equations this pathology is inevitable only in four
dimensions! Moreover, the space of the solutions within the
static spherically symmetric ansatz is richer comparing with the
four dimensional case.

In what follows, we shall find the general $D$-dimensional
solution within the ansatz (in sec.2) and will describe both its
metric properties (sec.3) and its hair (sec.4). We find many
black hole space-times with real hair.  The case $D=3$ we will
treat separately. We shall dicsuss what we have learned from
those concrete results in sec.5: Conclusions and Outlook.

%%%%%%%%%%%%%%%%%%%%%%%%%%%%%%%%%%%%%%%%%%%%%%%%%%%%%%%%%%%%%%%%%%%%
\section{Non-commutative action in {\it D} dimensions}
%%%%%%%%%%%%%%%%%%%%%%%%%%%%%%%%%%%%%%%%%%%%%%%%%%%%%%%%%%%%%%%%%%%
\setcounter{equation}{0}

Generalizing the 4-dimensional action obtained from
non-commutative geometry
to the arbitrary dimensional spacetimes (with metric of
signature $- +
\cdots +$) we get the new action\footnote{$V^\alpha$ is
spacelike field.}
\begin{equation}
I=\int_X d^Dx\sqrt{-g}\left[2R+4\kappa \nabla_\beta(\frac{V^\mu
V^\beta}
{\sqrt{V^2}})\nabla_\mu (\frac{1}{\sqrt{V^2}})\right],\label{newd}
\end{equation}
It what follows we
will find convenient to define the new fields $f^\alpha,~\sigma$
as $f^\alpha=\frac{V^\alpha}{\sqrt{V^2}}$, $\sigma=\frac{1}{\sqrt{V^2}}$
(similarily as in \cite{KKP1}).
Then the action becomes
\be
I=\int_X d^D x \sqrt{-g}\left[2R-4\kappa
f^{\alpha}f^{\beta}\frac{\nabla_
{\alpha}
\nabla_{\beta}\sigma}{\sigma}-4\kappa \Lambda(f^{\alpha}f{_\alpha}-1)
\right].\label{akcia}
\ee
with Lagrange multiplier $\Lambda$.
Variations of action (\ref{akcia}) with respect to $\Lambda$,
$f^\alpha$, $\sigma$, $g^{\alpha\beta}$ yield

\begin{eqnarray}
f^{\alpha}f^{\beta}g_{\alpha\beta}&=&1,\nonumber\\
f^{\alpha}\left(\frac{\nabla_{\mu}\nabla_{\alpha}\sigma}{\sigma}+
\Lambda g_{\mu\alpha}\right)&=&0, \label{matica}
\eea
\bea
\nabla_{\beta}\left(2f^{\alpha}f^{\beta}
\frac{\nabla_{\alpha}\sigma}{\sigma}-
\nabla_{\alpha}
(f^{\alpha}f^{\beta})\right)&=&0,\label{sig}
\end{eqnarray}
\begin{eqnarray}
R_{\mu\nu}-{\textstyle\frac12} g_{\mu\nu}R&=&\kappa\Bigg[-g_{\mu\nu}
\left(f^{\alpha}f^{\beta}\left(\frac{\nabla_{\alpha}
\nabla_{\beta}\sigma}
{\sigma}+
\Lambda g_{\alpha\beta}\right)-\Lambda\right)+
\nonumber\\
& &+2f^{\alpha}f_{~[\mu}\left(\frac{\nabla_{\alpha}
\nabla_{\nu]}\sigma}{\sigma}
+\frac{\nabla_{\alpha}\sigma \nabla_{\nu]}\sigma}{\sigma^2}
\right)+2f_\mu f_\nu\Lambda+
\nonumber\\
& &+\nabla^{\alpha}
\left(f_{\mu}f_{\nu}
\frac{\nabla_{\alpha}\sigma}{\sigma}\right)-2
\nabla_{\alpha}(f^{\alpha}f_{~[\nu})
\frac{\nabla_{\mu]}\sigma}{\sigma}\Bigg]. \label{einstrov}
\end{eqnarray}
where $[\alpha~\beta]$ means the symmetrization in the
indices, i.e.~$V_{[\alpha}V_{\beta]}=\frac{1}{2}(V_\alpha V_\beta
+V_\beta V_\alpha)$.
As we are interested in static, spherically symmetric solutions
of this  system, let
 our metric have the form
\be
ds^2=-e^{\nu(r)}dt^2+e^{\lambda(r)}dr^2+r^2d\Omega_{D-2}^2,
\label{dmetrika}
\ee
where $d\Omega_{D-2}^2$ is the standard round metric on the sphere
$S^{D-2}$.
In agreement with (\ref{matica}) we assume that only  $f^0$ and $f^1$
components of vector $f^\alpha$ are non-zero. In this ansatz the
 equations of motion are

\begin{eqnarray}
-e^\nu f^0f^{0}+e^\lambda f^1f^{1}&=&1,\label{vf}\\
\frac{\sigma'}{\sigma}\frac{\nu'}{2}e^{-\lambda}+\Lambda&=&0,
\label{2.9}\\
\frac {1}{ \sigma} \left(\sigma''-\frac{\lambda'}{2}\sigma'\right)+
e^\lambda\Lambda&=&0,\label{2.10}
\end{eqnarray}
\begin{equation}
{r^{D-2}} e^{\frac{\nu+\lambda}{2}}\left[
\left(f^1f^1\right)'+f^1f^1\left(\frac{\nu'}2+\lambda'-
\frac{2\sigma'}{\sigma}
+\frac {D-2} r \right)+f^0f^0\frac{\nu'}2 e^{\nu-\lambda}\right]=A
,\label{2.11}
\end{equation}

\begin{eqnarray}
& &\frac {(D-2)(D-3)} {2r^2}-(D-2)e^{-\lambda}\left(\frac {D-3}
{2r^2}-
\frac{\lambda'}{2r}
\right) =\kappa
\left(f^0f^0\right)'\frac{\sigma'}{\sigma}e^{\nu- \lambda}+
\nonumber\\
& &~~+\kappa\Bigg[f^0f^0\left(-\frac{\nu'}{2}
e^{\nu-\lambda}\frac{\sigma'}{\sigma}-\Lambda e^\nu\right)+
f^1f^1\left(\frac{\sigma''}{\sigma}-\frac{\lambda'}{2}
\frac{\sigma'}{\sigma}+\Lambda e^\lambda \right)-\Lambda\Bigg]+
\nonumber\\
& &
{}~~+\kappa f^0f^0e^{\nu-\lambda}\Bigg[\frac{\sigma'}{\sigma}
\Bigg(\frac 5 2 \nu'-\frac{\sigma'}{\sigma}-\frac{\lambda'}{2}
+\frac {D-2} r \Bigg)+\frac{\sigma''}{\sigma}+2\Lambda
e^\lambda\Bigg],\label{2.12}
\end{eqnarray}

\begin{eqnarray}
& &\frac {(D-2)(D-3)} {2r^2}
-(D-2)e^{-\lambda}\left(\frac{(D-3)}{2r^2}+
\frac{\nu'}{2r}
\right)=\kappa\left(f^1f^1\right)'\frac{\sigma'}{\sigma}+
\nonumber\\
& &~~+\kappa\Bigg[f^0f^0\left(-\frac{\nu'}{2}
e^{\nu-\lambda}\frac{\sigma'}{\sigma}-\Lambda e^\nu\right)+
f^1f^1\left(\frac{\sigma''}{\sigma}-\frac{\lambda'}{2}
\frac{\sigma'}{\sigma}+\Lambda e^\lambda
\right)-\Lambda\Bigg]+\nonumber\\
& &~~+
\kappa f^1f^1\Bigg[\frac{\sigma'}{\sigma}
\Bigg(-\frac{\sigma'}{\sigma}
+\frac{5}{2} \lambda'+
\frac{\nu'}{2}+
\frac{(D-2)}{r}
\Bigg)-3\frac{\sigma''}{\sigma}-
2e^\lambda\Lambda\Bigg]+\nonumber\\
& &~~+\kappa f^0f^0\nu'e^{\nu-\lambda}\frac{\sigma'}{\sigma},
\label{2.13}
\end{eqnarray}

\begin{eqnarray}
\frac{D-4}{4r}e^{-\lambda}\left(\nu'-\lambda'+\frac{2}{r}(D-3)
(1-e^{\lambda})\right)+
\frac{e^{-\lambda}} 4
\Big(2\nu''-\nu'(\lambda'-\nu')+\nonumber\\
+\frac{D-2}{r}(\nu'-\lambda')\Big)=
\kappa\Lambda,
\end{eqnarray}
(where (\ref{2.11}) is actually the first integral of
(\ref{sig}) with an integration constant $A$).

%%%%%%%%%%%%%%%%%%%%%%%%%%%%%%%%%%%%%%%%%%%%%%%%%%%%%%%%%%%%%%%%
\section{The solution of the field equations in $D$ dimensions}
%%%%%%%%%%%%%%%%%%%%%%%%%%%%%%%%%%%%%%%%%%%%%%%%%%%%%%%%%%%%%%%%%%
\setcounter{equation}{0}

Extracting $\Lambda$ from Eq.~(\ref{vf}) and
Eq.~(\ref{2.10})\footnote{We assume
$\sigma'\not=0$, because otherwise we would end up with standard
Schwarzschild case (see \cite{KKP1}).} and integrating the obtained
relation we (as in \cite{KKP1}) arrive at
\begin{equation}
e^\nu =\sigma'^2 e^{-\lambda} e^{-2K},\label{ni}
\end{equation}
where $K$ is a constant.
Substituting Eq.~(\ref{ni}) and
Eq.~(\ref{vf}) into Eq.~(\ref{2.11})
we obtain

\begin{equation}
\left(f^1f^1\right)'+2f^1f^1\left(\frac{D-2}{2r}+
\frac{\sigma''}{\sigma'}
-\frac{\sigma'}\sigma\right)=\frac{A\,e^K}{r^{D-2}\sigma'}
+e^{-\lambda}
\left(\frac{\sigma''}{\sigma'}-\frac{\lambda'}2\right).
\label{nnewf12}
\end{equation}
The difference of Eq.~(\ref{2.12}) and Eq.~(\ref{2.13})
results in

\begin{equation}
\frac{\sigma''}\sigma +\kappa\left(\frac{\sigma'^2}{\sigma^2}-
\frac{r}{D-2}
\frac{\sigma'^3}{\sigma^3}\right)=0,\label{nsigma}
\end{equation}
and Eq.~(\ref{2.13}), combined with Eq.~(\ref{nnewf12}), yields

\begin{eqnarray}
 f^1f^1 &=&\frac{(D-2)(D-3)\sigma^2}{2\kappa r^2\sigma'^2}-
\frac{(D-2)e^{-\lambda}}{2\kappa}\Bigg(\frac
{(D-3)\sigma^2}{r^2\sigma'^2}+\frac{2\sigma^2\sigma''}{r\sigma'^3}
-\nonumber\\
& &-\frac
{\lambda'\sigma^2}{r\sigma'^2}\Bigg)-
\frac{A\sigma e^K}{\sigma'^2
r^{D-2}}.\label{newf12}
\end{eqnarray}
Using Eq.~(\ref{vf}) and inserting $f^1f^1$ from
Eq.~(\ref{newf12}) into
Eq.~(\ref{sig}), we obtain

\begin{equation}
\frac{\sigma^2}{\sigma'^2}Q''+\left(\kappa\frac\sigma{\sigma'}-
\frac{\sigma''\sigma^2}{\sigma'^3}\right)Q'-
\kappa\frac{D-3}{D-2}Q-
(D-3)(D-4)e^{-2K}\sigma^2r^{D-5}
=0,\label{nkve}
\end{equation}
where we have introduced $Q(r)$, defined by
$$Q(r)\equiv e^{\nu(r)}r^{D-3}.$$
The equation (\ref{nsigma}) can be easily solved
after changing variable $r\to\sigma$, as {\it then} it becomes
\begin{eqnarray}
\frac{d^2 r}{d\sigma^2}-\frac{\kappa}\sigma\frac{dr}{d\sigma}+
\frac{\kappa r}
{(D-2)\sigma^2}&=&0.\label{nforr}
\end{eqnarray}
The general solution is
\begin{eqnarray}
r&=&c_1\sigma^{\alpha_1}+c_2\sigma^{\alpha_2},\label{nr}
\end{eqnarray}
where $$\alpha_{1,2}=\frac 1 2\Bigg(1+\kappa\pm
\sqrt{1+2\kappa\frac
{D-4}{D-2}+\kappa^2}\Bigg). $$
It is straightforward to solve the Eq.~(\ref{nkve})
in $D=3$.
The general solution in this case is

\begin{eqnarray}
Q(\sigma)&=& k_1\ln\sigma +k_2  ~~~~\kappa=1,
\label{3Q}\\
Q(\sigma)&=& k_1\frac {\sigma^{1-\kappa}}
{1-\kappa}+k_2 ~~~ \kappa \not= 1.
\end{eqnarray}
In higher dimensions we change the variable $r\to\sigma$, and
obtain

\begin{eqnarray}
\frac{d^2 Q}{d\sigma^2}+\frac{\kappa}\sigma\frac{dQ}{d\sigma}-
\frac{\kappa (D-3) Q}
{(D-2)\sigma^2}=(D-3)(D-4)e^{-2K}r^{D-5}.\label{nforQ}
\end{eqnarray}
This equation is the Euler's differential equation with non-zero
right-hand-side.
The general solution of Eq.~(\ref{nforQ}) is given by
\begin{eqnarray}
Q(\sigma)&=&k_1\sigma^{\beta_1}+k_2\sigma^{\beta_2}
+\sum_{j=0}^{D-5}A_j \sigma^{\omega_j}, \label{nQ}
\end{eqnarray}
where
 $$\beta_{1,2}=\frac 1 2 \Bigg(1-\kappa\pm \sqrt{1+2\kappa\frac
{D-4}{D-2}+\kappa^2}\Bigg),$$
$$\omega_j=j(\alpha_2-\alpha_1)+(D-5)\alpha_1+ 2,$$
$$A_j=\frac{{D-5 \choose  j}(D-3)(D-4)c_1^{D-5-j}c_2^{j}e^{-2K}}
{\omega_j(\omega_j -1)+\kappa(\omega_j-\frac{D-3}{D-2})},$$
$c_{1,2}$ and $k_{1,2}$ are (real) constants, and
$j=0,1,2,\dots, D-5$.

%%%%%%%%%%%%%%%%%%%%%%%%%%%%%%%%%%%%%%%%%%%%%%%%%%%%%%%%%%%%%%%%%%%
\section{The analysis of the solution in $D$ dim.}
%%%%%%%%%%%%%%%%%%%%%%%%%%%%%%%%%%%%%%%%%%%%%%%%%%%%%%%%%%%%%%%%%%
\setcounter{equation}{0}

At this moment we have only ``raw" solutions of equations of
motion.
In principle, we know everything about the system.
 From (\ref{nr}) we can obtain
$\sigma$. Formulae (\ref{nQ}) and (\ref{ni}) give us
the metric of the spacetime.
$V^0$ and $V^1$ can be obtained using (\ref{newf12}) and the
definitions of
$f^{0},~f^{1}$.
Now we shall study
deeper the properties of the solutions. Let us start with
3-dimensional case.

%%%%%%%%%%%%%%%%%%%%%%%%%%%%%%%%%%%%%%%%%%%%%%%%%%%%%%%%%%%%%%%%%%
\subsection{Black-hole metric and the scalar curvature in
3-dim.}
%%%%%%%%%%%%%%%%%%%%%%%%%%%%%%%%%%%%%%%%%%%%%%%%%%%%%%%%%%%%%%%%%
\setcounter{equation}{0}

The properties of the obtained spacetime vary with the choice of
the integration constants. In what follows, we shall consider
only
the cases which are the most obvious candidates for the black
hole
metric.

{Case} I.~Let $c_2=0$ and $\kappa>1$. Then
$\sigma=\frac{r}{c_1}$ and
\begin{equation}
e^{\nu(r)}= k_1 c_1^{\kappa-1}\frac
{r^{1-\kappa}}{1-\kappa}+k_2,~~
{}~~e^{\lambda(r)}=\frac{e^{-2K}c_1^{-2}}{k_1 c_1^{\kappa-1}\frac
{r^{1-\kappa}}
{1-\kappa}+k_2}. \label{jedna}
\end{equation}
It is easy to see that $e^\nu\to
k_2,~e^\lambda\to\frac{e^{-2K}}{c_1^2 k_2}$
for $r\to\infty$. We know that $\sigma$ is positive, thus
$c_1>0$.
In the asymptotic region ($r\to\infty$) we need
$g_{00}<0$
and, therefore $k_2>0$. Then
$g_{11}>0$ for $r\to\infty$, as it should. Moreover, there is an
event horizon if $k_1>0$.

{Case} ${\rm I\!I}$.~Let $c_1=0$ and $0<\kappa<1$. Then
$\sigma(r)=c_2^{-\frac{1}{\kappa}}r^{\frac{1}{\kappa}}$,
\begin{equation}
e^{\nu(r)}=k_1 c_2^{\frac{\kappa-1}{\kappa}}\frac
{r^{\frac{1-\kappa}{\kappa}}}
{1-\kappa}+k_2,~~~~
e^{\lambda(r)}=\frac{e^{-2K}\kappa^{-2}
c_2^{-\frac{2}{\kappa}}r^{\frac{2-2\kappa}{\kappa}} }
{\frac{k_1}{1-\kappa}
c_2^{\frac{\kappa-1}{\kappa}}r^{\frac{1-\kappa}\kappa}
+k_2}. \label{dva}
\end{equation}
{}From $\sigma>0$ it follows $c_2>0$ and from the asymptotic
behavior of
$g_{00}$ we have
$k_1>0$. Then the horizon does exist if $k_2<0$ and it is given by
\be
r_H=c_2\left(-\frac{k_2(1-\kappa)}{k_1}\right)^{\frac{\kappa}
{1-\kappa}}.
\ee
Is there any curvature singularity hidden behind the horizon?
The formula for scalar curvature in $D$-dimensional spacetime
with metric of form (\ref{dmetrika}) is
\begin{eqnarray}
R&=&\frac{(D-2)(D-3)}{r^2}-e^{-\lambda}\Bigg(\frac{(D-2)(D-3)}
{r^2}+
\frac{\nu'^2}{2}+\nu''+\nonumber\\
& &+(D-2)\frac{\nu'-\lambda'}{r}-
\frac{\nu'\lambda'}{2}\Bigg). \label{curva}
\end{eqnarray}
Using this formula for $D=3$
 we can calculate that $R\to\infty$ for
$r\to 0$, $R\to 0$ for $r\to\infty$ and $R$ is finite for
$r\not=0$.
Thus in both
``right" cases (I, $\rm I\!I$) the spacetime
is asymptotically flat, smooth at the horizon and
singular at $r=0$\footnote{In fact, we have checked the
statement
by computing all components of the Riemann tensor in parallelly
propagated orthonormal vielbein. \label{footn}}.

Now we are going to analyze the behavior of hair.

%%%%%%%%%%%%%%%%%%%%%%%%%%%%%%%%%%%%%%%%%%%%%%%%%%%%%%%%%%%%%%%%%%
\subsection{Hair in 3-dim.}
%%%%%%%%%%%%%%%%%%%%%%%%%%%%%%%%%%%%%%%%%%%%%%%%%%%%%%%%%%%%%%%%%

Let us first study the case I when $c_2=0$ and $\kappa>1$.
Inserting the metric (\ref{jedna}) into Eq.~(\ref{newf12}) for
$D=3$ we obtain a strange kind of singularity. Our $f^1,~f^0$
are real and it implies that $f^1f^1$ and $f^0f^0$ have always
to be positive. But for our choice of integration constants,
$f^1f^1$ cannot be positive in all spacetime. Nevertheless,
demanding $A>\frac{1-\kappa}{2\kappa}c_1k_2e^K$ we can arrange
that the region where $f^1f^1$ is negative is hidden behind the
horizon.  Then if we require the asymptotical positivity of
$f^0f^0$ , we will obtain another condition, viz.~$A<c_1k_2e^K$,
which has to be also satisfied. It is clear that both
inequalities cannot be satisfied at the same time for watever
$\kappa>0$.  Therefore this case is out of our interest.

The next (and in $D=3$ the most interesting) case is when $c_1=0$ and
$0<\kappa<1$. Inserting the metric (\ref{dva}) into the
(\ref{newf12}) we obtain
$$
f^1f^1=-e^K\kappa^2c_2^{\frac 1
\kappa}(e^K\frac{\kappa}{2}c_2^{\frac 1 \kappa}+A)r^{1-\frac 1
\kappa}.
$$
Using the definition $f^1=\sigma V^1$ we have
\begin{equation}
V^1V^1=-e^K\kappa^2c_2^{\frac 3
\kappa}(e^K\frac{\kappa}{2}c_2^{\frac 1 \kappa}+A)r^{1-\frac 3
\kappa}. \label{hair1}
\end{equation}
Let
\be
A\leq-\frac 1 2 \kappa e^K c_2^{\frac 1 \kappa}. \label{poda}
\ee
{}From equation (\ref{hair1}) it is then clear that $V^1V^1\geq0$
everywhere, in all spacetime and that $V^1V^1\to\infty$ when
$r\to 0$. Using Eq.~(\ref{vf}), we can see that
\begin{equation}
V^0V^0=-g_{11}g^{00}V^1V^1+\frac
{g^{00}}{\sigma^2}, \label{hair2}
\end{equation}
so that $V^0V^0$ is positive below the
horizon. Because $V^0V^0\sim -\frac{c_2^{\frac 2 \kappa}}{k_2}
r^{-\frac{2}{\kappa}}$ for $r\to 0$, it is clear that
$V^0V^0\to\infty$ for $r\to 0$.
Can
$V^0V^0$ be negative somewhere
above the horizon? Suppose that $V^0V^0=0$ in some $r=r_{crit}$,
i.e.
\be
Yr_{crit}^{\frac{2}{\kappa}-2}
Zr_{crit}^{1-\frac{1}{\kappa}}
e^{-2\nu(r_{crit})}-e^{-\nu(r_{crit})}=0, \label{crit}
\ee
where we have introduced the costants $Y,~Z$ as follows
$$
e^{\nu(r)}=
Yr^{\frac{1-\kappa}{\kappa}}+k_2,~{\rm and}
{}~~f^1f^1=Zr^{1-\frac 1 \kappa}
$$
$$
Y=k_1 c_2^{\frac{\kappa-1}{\kappa}}\frac{1}{1-\kappa},
{}~~~Z=-e^K\kappa^2c_2^{\frac 1
\kappa}(e^K\frac{\kappa}{2}c_2^{\frac 1 \kappa}+A).
$$
Inserting the metric (\ref{dva}) into the Eq.~(\ref{crit}), we
obtain
$$
-k_2=(X-YZ)r_{crit}^{\frac 1 \kappa-1},
$$
where
$$
e^{\lambda(r)}=
Xr^{\frac{2-2\kappa}{\kappa}} e^{-\nu},~~~
{\rm i.e.}~ X=e^{-2K}\kappa^{-2}
c_2^{-\frac{2}{\kappa}}.
$$
Let
\be
A<-\frac 1 2 \kappa e^K c_2^{\frac 1 \kappa}-
\frac{e^{-K}c_2k_1}{1-\kappa}.\label{podb}
\ee
Calculating then the coefficient $(X-YZ)$, we
see that it is negative and this implies that $r_{crit}$ does not
exist, because $k_2$ is also negative. This means that
the $V^0V^0$ does not change its sign above the horizon.
The sign is plus as it is seen from the asymptotical
behavior of  $V^0V^0$.
Note also that only one condition for $A$ is to be fulfilled because
 (\ref{podb}) implies (\ref{poda}).

We conclude that in the case ${\rm I\!I}$ the hair is real
everywhere.
Recall that in four dimensions there was no such solution
\cite{KKP1}.

%%%%%%%%%%%%%%%%%%%%%%%%%%%%%%%%%%%%%%%%%%%%%%%%%%%%%%%%%%%%%%%%
\subsection{Black-hole metric and the scalar curvature in
dimensions $D\ge 5$}
%%%%%%%%%%%%%%%%%%%%%%%%%%%%%%%%%%%%%%%%%%%%%%%%%%%%%%%%%%%%%%%%

Consider $c_1=0$.
Then
$\sigma=\left(\frac{r}{c_2}\right)^{\frac{1}{\alpha_2}}$
and the metric

\be
e^\nu=\left[A_{D-5}\left(\frac{r}{c_2}\right)^{D-5+\frac{2}
{\alpha_2}}
+k_1\left(\frac{r}{c_2}\right)^{\frac{\beta_1}{\alpha_2}}+
k_2\left(\frac{r}{c_2}\right)^{\frac{\beta_2}{\alpha_2}}
\right]
r^{3-D},\label{enani}
\ee
and, consequently
\be
e^\lambda=\left[A_{D-5}\left(\frac{r}{c_2}\right)^{D-5+\frac{2}
{\alpha_2}}
+k_1\left(\frac{r}{c_2}\right)^{\frac{\beta_1}{\alpha_2}}+
k_2\left(\frac{r}{c_2}\right)^{\frac{\beta_2}{\alpha_2}}
\right]^{-1} \frac{r^{D-5}}{e^{2K}\alpha_2^2}
\left(\frac{r}{c_2}\right)^{\frac{2}{\alpha_2}}. \label{enalamb}
\ee
Note that $A_{D-5}>0$.
We would like to find the black hole solution.
Let us set one of
the integration constants $k_1,~k_2$ equal to zero. Start with the
case
when $k_1=0$.
Then the horizon does exist if $k_2<0$:
$$
r_H=c_2 \left(-\frac{A_{D-5}}{k_2}\right)^{\frac{\alpha_2}{\beta_2-
(D-5)\alpha_2-2}}.
$$
The another case is when $k_2=0$. Then the horizon is located at
$$
r_H=c_2 \left(-\frac{A_{D-5}}{k_1}\right)^{\frac{\alpha_2}{\beta_1-
(D-5)\alpha_2-2}}.
$$

The direct computation using (\ref{curva}) shows in both cases
that $R$ diverges
for $r\to 0$,
tends to zero for $r\to\infty$ and is finite for
$r\not=0$\footnote{Footnote \ref{footn} from page 9 applies here
too.}.
The conclusion is that our horizons hide the curvature
singularities, i.e.~we have the standard
black-holes.

%%%%%%%%%%%%%%%%%%%%%%%%%%%%%%%%%%%%%%%%%%%%%%%%%%%%%%%%%%%%%%%%%%
\subsection{The hair in $D\geq 5$}
%%%%%%%%%%%%%%%%%%%%%%%%%%%%%%%%%%%%%%%%%%%%%%%%%%%%%%%%%%%%%%%%%%

First of all, let us study the behavior of
$V^2=V^\alpha V_{\alpha}$, the only scalar which can be
constructed
from our vector field $V^\alpha$. Because we are interested in
the
case when $c_1=0$ we have
$$
V^2= \left(\frac{r}{c_2}\right)^{-\frac{2}{\alpha_2}}.
$$
{}From this  expression it is clear that $V^2$
is positive everywhere, tends to  zero for $r\to\infty$ and is
{\it smooth} and
{\it bounded} at the horizon.

Let us go to study the components of vector $V^\alpha$.
Let $k_1=0$.
Inserting the metric (\ref{enani}), (\ref{enalamb}) into
(\ref{newf12}) we can see the behavior of $f^1f^1$.
$f^1f^1\to {\it finite~ positive~ constant}$
and $f^1f^1\to -\infty$ when
$r\to\infty$ and $r\to 0$ respectively. There is one null point
of $f^1f^1$ in $r=r_{crit}$, which depends on
integration constants. Choosing the appropriate integration
constant $A$, we can move this null point below the horizon.
Then $f^1f^1>0$ for $r>r_{crit}$.
The analysis of $f^0f^0$ shows that $f^0f^0\to\infty$ and
$f^0f^0\to 0$ for $r\to 0$ and $r\to\infty$ respectively, but
it is not necessarily positive everywhere.
To ensure the positivity of $f^0f^0$ at least for
$r>r_{crit}$ (while $r_{crit}<r_H$) we have to satisfy the following
 condition (cf.~the similar case in $D=4$ \cite{KKP1})
$$
\kappa\leq1+\frac{2}{D-4}.
$$
The behavior of $V^0V^0$ is qualitatively the same as the
behavior of $f^0f^0$. Note that in this case of choice of
integration
constant the behavior of hair is very similar to 4-dimensional
case (see \cite{KKP1}).

Qualitatively new
(and in $D\ge 5$  the most interesting) case occurs for $k_2=0$.
Then the metric has the form
 $$
e^\nu=A_{D-5}\left(\frac{1}{c_2}\right)^{D-5+\frac{2}{\alpha_2}}
r^{-2+\frac{2}{\alpha_2}}
+k_1\left(\frac{1}{c_2}\right)^{\frac{\beta_1}{\alpha_2}}
r^{\frac{\beta_1}{\alpha_2}-(D-3)}, \label{enanidob}
$$
$$
e^\lambda=\left[A_{D-5}\left(\frac{r}{c_2}\right)^{D-5+\frac{2}
{\alpha_2}}
+k_1\left(\frac{r}{c_2}\right)^{\frac{\beta_1}{\alpha_2}}
\right]^{-1} \frac{r^{D-5}}{e^{2K}\alpha_2^2}
\left(\frac{r}{c_2}\right)^{\frac{2}{\alpha_2}}.
\label{enalambdob}
$$
Inserting this metric into the expression

\begin{eqnarray}
 V^1V^1 &=&\frac{(D-2)(D-3)}{2\kappa r^2\sigma'^2}-
\frac{(D-2)e^{-\lambda}}{2\kappa}\Bigg(\frac
{(D-3)}{r^2\sigma'^2}+\frac{2\sigma''}{r\sigma'^3}-\nonumber\\
& &-\frac
{\lambda'}{r\sigma'^2}\Bigg)-
\frac{Ae^K}{\sigma\sigma'^2
r^{D-2}},
\end{eqnarray}
we obtain that

\be
V^1V^1=\frac{D-2}{2\kappa\sigma'^2}\Bigg[Cr^{-2}+\Bigg(-Ac_2^
{\frac{1}{\alpha_2}}
\frac{2\kappa e^K}{D-2}+E\Bigg)r^{-(D-2)-\frac{1}{\alpha_2}}\Bigg],
\label{v1}
\ee
where
$$
C=D-1-\frac 1 {\alpha_2}-A_{D-5}(D-3)e^{2K}\alpha_2^2c_2^{5-D}
$$
and
$$
E=k_1\alpha_2^2(-\frac 1
{\alpha_2}+1)e^{2K}c_2^{\frac{2-\beta_2}{\alpha_2}}.
$$
It can be shown that $C$ is a positive constant.
{}From (\ref{v1}) it is clear that using appropriate constant $A$,
namely
 $A<c_2^{-\frac{1}{\alpha_2}}
\frac{D-2}{2\kappa e^K} E $, our
hair $V^1V^1$ is positive in all spacetime.
What about the behavior of $V^0V^0$?
 From the formula (\ref{hair2}) it follows  that
$V^0V^0$ is positive below the horizon, and using (\ref{v1}), it
can be
shown that $V^0V^0$ is positive also everywhere above the
horizon.
Thus this case the hair $V^0V^0$ and $V^1V^1$ are positive
everywhere.

Note that $V^0V^0$ diverges at the horizon, though the invariant
$V^\alpha V_\alpha$ is smooth everywhere, including the horizon.
It is not clear, however, how pathological is the divergent
behavior of the non-invariant components of the field. It may be,
that other propagating field coupled to $V^\alpha$ do not feel any
singularity at the horizon (as it is the case of another well-known
case with the singular hair, i.e.~the Bekenstein black
hole\footnote
{In that case, however, even a truly invariant quantity was
singular at the horizon.}
\cite{bek2}). The problem certainly requires a deeper analysis.

%%%%%%%%%%%%%%%%%%%%%%%%%%%%%%%%%%%%%%%%%%%%%%%%%%%%%%%%%%%%%%%%%%%
\section{Conclusions and outlook}
%%%%%%%%%%%%%%%%%%%%%%%%%%%%%%%%%%%%%%%%%%%%%%%%%%%%%%%%%%%%%%%%%%%
\setcounter{equation}{0}

We have obtained the general $D$-dimensional static spherically
symmetric solutions of the specific vector $\sigma$-model coupled
to the Einstein gravity. The model arises in the studies of pure
gravity in the non-commutative geometry setting. We have found a
large subclass of the solutions having the structure of standard
assymtotically flat black holes with a smooth horizon covering the
curvature singularity. We have interpreted the ``non-commutative"
components of the metric as the hair and we found the dependence of
its properties on the choice of integration constants. The latter
turned out to be nontrivial and, in fact, quite restrictive. Unlike
the case of $D=4$ \cite{KKP1} (where the hair was necessarily
imaginary near the singularity), there are the black hole solutions
with the real hair everywhere! Thus, the peculiar singularity of
the four dimensional black hole space-times can be removed by
increasing the dimension.

We believe that the model (\ref{newd}) deserves further study,
because maintaining
the reparametrization invariance it decreases the propagating
degrees of freedom of the standard gravity. Indeed, it can be
seen from Eq.~(\ref{matica}), that the ``direction" fields
$f^\alpha$ play essentially the role of the Lagrange multiplier.
This property may be very important for further quantization. It
can be also very interesting to find the consequences of the fact
that the theory has the geometrical origin in the sense of
non-commutative geometry.

\end{document}